\input ./style/arxiv-general.cfg
\documentclass[MSNbibl,nameyear,dvips]{arxstspdf}
\makeatletter
   \@ifpackageloaded{graphicx}{}{\usepackage{graphicx}}
\makeatother
\usepackage{flushend}
\usepackage{stfloats}
\volume{30}
\issue{2}
\pubyear{2015}
\firstpage{228}
\lastpage{241}
\doi{10.1214/14-STS508}
\docsubty{FLA}

\makeatletter
\newcommand{\ds}{\displaystyle}

\newtheorem{theorem}{Theorem}
\newtheorem{lemma}{Lemma}
\newproclaim{defn}{Definition}
\makeatother

\begin{document}
\begin{frontmatter}

\title{Posterior Model Consistency in Variable Selection as the Model
Dimension Grows}
\runtitle{Posterior model consistency}

\begin{aug}
\author[A]{\fnms{El\'{\i}as}~\snm{Moreno}\corref{}\ead[label=e1]{emoreno@ugr.es}},
\author[B]{\fnms{Javier}~\snm{Gir\'{o}n}\ead[label=e2]{fj\_giron@uma.es}}
 \and
\author[C]{\fnms{George}~\snm{Casella}}
\runauthor{E. Moreno, J. Gir\'{o}n and G. Casella}

\affiliation{University of Granada, University of M\'{a}laga and  University of Florida}

\address[A]{El\'{\i}as Moreno is  Professor of Statistics, Department of Statistics,
University of Granada, 18071 Granada, Spain \printead{e1}.}
\address[B]{Javier Gir\'{o}n is Professor of Statistics,
Departament of Statistics, University of M\'alaga,  29016 M\'alaga,  Spain \printead{e2}.}
\address[C]{George Casella was  Distinguished Professor,
University of Florida, Florida, USA, deceased, June 17, 2012.}
\end{aug}

\begin{abstract}
Most of the consistency analyses of Bayesian procedures for variable
selection in regression refer to pairwise consistency, that is, consistency
of Bayes factors. However, variable selection in regression is carried out
in a given class of regression models where a natural variable selector is
the posterior probability of the models.

In this paper we analyze the consistency of the posterior model
probabilities when the number of potential regressors grows as the sample
size grows. The novelty in the posterior model consistency is that it
depends not only on the priors for the model parameters through the Bayes
factor, but also on the model priors, so that it is a useful tool for
choosing priors for both models and model parameters.

We have found that some classes of priors typically used in variable
selection yield posterior model inconsistency, while mixtures of these
priors improve this undesirable behavior.

For moderate sample sizes, we evaluate Bayesian pairwise variable selection
procedures by comparing their frequentist Type I and II error probabilities.
This provides valuable information to discriminate between the priors for
the model parameters commonly used for variable selection.
\end{abstract}

\end{frontmatter}

\section{Introduction}\label{sec1}

In some applications of regression models to complex problems, for instance,
in genomic, clustering, change points detection, etc., the dimension of the
parameter space of the sampling models is either very large or grows with
the sample size. The question we address here is whether consistency of the
Bayesian variable selection approach still holds in this setting. A~partial
answer to this question was given in  \citet{MorGirCas10}, where consistency
of the Bayes factors (pairwise consistency) when the number of regressors $k$
increases with rate $k=O(n^{b})$, $b\leq 1$, was considered. It was there
proved that any pair of nested regression models for which the Bayes factor
has an asymptotic approximation equivalent to the BIC \citep{Sch78} is
consistent for $b<1$ but it is not for $b=1$. Note that the BIC is a valid
approximation for a wide class of prior distributions on the model
parameters. It was also seen that the Bayes factor for the intrinsic priors
considerably improves the BIC behavior for small or moderate sample sizes
\citep{Casetal09}.

Nevertheless, variable selection in regression is a model selection problem
in a class $\mathfrak{M}$ of $2^{k}$ normal regression models, and we wonder
if the Bayes factor consistency when $k=O(n^{b})$, $b\leq 1$, can be
extended to posterior model consistency in the class of models~$\mathfrak{M}$.
The use of the posterior model probabilities as a variable selector
procedure implies that variable selection is understood as a decision
problem where the decision space $\mathfrak{D}$ and the space of states of
nature $\mathfrak{M}$ are the same. Assuming a $0\mbox{--}1$ loss function on the
product space $\mathfrak{D\times M}$, the optimal decision is that of
choosing the model with the highest posterior probability; other loss
functions can indeed be used; see, for instance, the review paper by \citet{ClyGeo04}.

Posterior model consistency in $\mathfrak{M}$ is understood as the
convergence to one, in probability, of the sequence of the posterior
probabilities of the true model. We are considering the true model to be the
one from which the observations are drawn. We note that the frequentist and
Bayesian consistency notions do not necessarily coincide. For instance, \citet{Sha97} defines a true model to be the submodel minimizing the average
squared prediction error, and consistency of a model selection procedure
means that the selected model converges in probability to this model.

From the necessary and sufficient conditions we give to achieve posterior
model consistency it follows that Bayes factor consistency does not
necessarily yield posterior model consistency. This was already pointed out
by \citet{JohRos12}. Further, posterior model consistency of a
Bayesian procedure in $\mathfrak{M}$ depends on the Bayes factor, the prior
over the class of models $\mathfrak{M}$ and the rate of growth of $k$, and
thus it has to be studied in a case-by-case basis.

The Bayes factors we review here are those obtained using the intrinsic
priors on the model parameters (\cite{BerPer96};  \cite{Mor97}; \cite{MorBerRac98}) and a couple of versions of the Zellner's $g$-priors
(\cite{ZelSio80}; \cite{Zel86}). These versions include the $g$-priors
with $g=n$ and the prior obtained as a mixture of $g$-priors with respect
to the $\operatorname{InverseGamma}(g|1/2,n/2)$.
This latter prior was recommended by
\citet{ZelSio80} and considered in \citet{Liaetal08} and \citet{ScoBer10}, among others. As we will see, these Bayes factors exhibit
different dimension corrections, that suggest a different behavior for
moderate sample sizes, a point that we also explore here.

The priors over the set of models we review are the independent Bernoulli
parametric class $\{\pi (M|\theta )$, $0<$ $\theta <1\}$ introduced by
\citet{GeoMcC93} and a specific mixture of these priors which we
refer to as the hierarchical uniform model prior $\pi^{\mathrm{HU}}(M)$. This latter
prior is a particular case of a set of hierarchically uniform priors
considered by \citet{GeoMcC93}, who argued that ``\textit{one may wish to weight more according the model
size}.''

Related posterior model consistency for variable selection for homoscedastic
high-dimensional regression models was analyzed by \citet{JohRos12}.
They considered Bayes factors for nonlocal priors on the regression
parameters, an inverse gamma for the common variance errors, and models
priors such that $\pi (M_{t})/\pi (M)$ $>\varepsilon >0$ for any $M\in
\mathfrak{M}$, where $M_{t}$, the true model, is a fixed model. We note that
the Bernoulli class of model priors and the hierarchical uniform model prior
$\pi^{\mathrm{HU}}(M)$ are excluded from their analysis. Further, the rate of growth
of the number of regressors does not play a relevant role for the posterior
model consistency of their Bayesian models, while for the Bayesian models
considered here it does.

\subsection{Notation}\label{sec11}

Let $Y$ represent an observable random variable and $X_{1},\ldots,X_{k}$ a
potential set of explanatory regressors related through the normal linear
model
\[
Y=\beta _{0}+\beta _{1}X_{1}+\cdots+\beta
_{k}X_{k}+\varepsilon _{k},\quad \varepsilon
_{k}\sim N\bigl(0,\sigma _{k}^{2}\bigr),
\]
where the vector of regression coefficients $\bolds{\beta}_{k+1}=(\beta_{0},\beta _{1},\ldots,\beta _{k})^{\prime }$ and the variance error $\sigma
_{k}^{2}$ are unknown. Let $(\mathbf{y},\mathbf{X})$ be the data set,
where $\mathbf{y}$ is a vector of $n$ independent observations of $Y$ and
$\mathbf{X}$ a $n\times (k+1)$ design matrix of full rank. This full sampling normal
model $N_{n}(\mathbf{y}|\mathbf{X}\bolds{\beta}_{k+1},\sigma _{k}^{2}\mathbf{I}_{n}) $ is denoted as
$M_{k}$ and the simplest intercept only normal model
$N_{n}(\mathbf{y}|\beta_{0}\mathbf{1}_{n}, \sigma_{0}^{2}\mathbf{I}_{n})$
as $M_{0}$. We remark that the regression coefficients change across models,
although for simplicity we use the same alphabetical notation.

It is convenient to split the class $\mathfrak{M}$ of regression models
involved in variable selection as follows. By $\mathfrak{M}_{j}$ we denote
the class of models with $j$ regressors, $0\leq j\leq k$, the number of
which is\vspace*{1pt} ${k \choose j}$, and by $M_{j}$ we denote a generic model in
$\mathfrak{M}_{j}$ with sampling density $N_{n}(\mathbf{y}|\mathbf{X}_{j+1}
\bolds{\beta}_{j+1}, \sigma_{j}^{2}\mathbf{I}_{n})$, where
$\bolds{\beta}_{j+1}=(\beta _{0},\beta _{1},\ldots,\beta _{j})^{\prime}$ is the
unknown\vspace*{1pt} vector of regression coefficients, $\mathbf{X}_{j+1}$ a $n\times
(j+1)$ submatrix of $\mathbf{X}$ and $\sigma _{j}^{2}$ the unknown variance
error. Therefore, $\mathfrak{M}=\bigcup _{j=0}^{k}\mathfrak{M}_{j}$. The
developments in the paper will be clear using this somewhat ambiguous, but
simpler, notation.

\subsection{Summary}\label{sec12}

We find that when $k$ grows with $n$, the intrinsic priors for model
parameters are preferred to either the $g$-prior for $g=n$ or the mixtures of
$g$-priors, and the hierarchical uniform model prior is preferred to the
Bernoulli model prior for any fixed value of the hyperparameter $\theta \in
(0,1)$.

The rest of the paper is organized as follows. In Section~\ref{sec2} we give
necessary and sufficient conditions to achieve posterior model consistency.
In Section~\ref{sec3}\vadjust{\goodbreak} we give asymptotic approximations to the Bayes factors for the
$g$-priors with $g=n$, for the mixture of $g$-priors and for the intrinsic
priors over the model parameters, for $k=O(n^{b})$, $0\leq b\leq 1$. In
Section~\ref{sec4} posterior model consistency for the Bayesian procedures is
presented. Section~\ref{sec5} contains a sampling evaluation of the three Bayes
factors for moderate sample sizes. A summary of the conclusions is given in
Section~\ref{sec6}, and the  \hyperref[appA]{Appendix} contains the proofs of most of the results.

\section{Posterior Model Consistency}\label{sec2}

Given a data set $(\mathbf{y},\mathbf{X})$ coming from a linear model in
$\mathfrak{M}$, and the priors for the models and model parameters
$\{\pi (\bolds{\beta}_{j+1},\sigma _{j}|M_{j}),\pi (M_{j})$, $M_{j}\in
\mathfrak{M}_{j}$, $j=0,1,\ldots,k\}$, the posterior probability of a generic
model $M_{j}$ can be written as
%
\begin{equation}
\label{equ1}\quad \Pr (M_{j}|\mathbf{y},\mathbf{X})=\frac{B_{j0}(\mathbf{y},\mathbf{X})
\pi (M_{j})}{\sum_{i=0}^{k}\sum_{M_{i}\in \mathfrak{M}_{i}}B_{i0}(\mathbf{y},
\mathbf{X})\pi (M_{i})},\hspace*{-6pt}
\end{equation}%
where $B_{j0}(\mathbf{y},\mathbf{X})$ denotes the Bayes factor for comparing
models $M_{j}$ and $M_{0}$, which is given by
\begin{eqnarray*}
&& B_{j0}(\mathbf{y},\mathbf{X})\\
&& \quad=\biggl(\int N_{n}\bigl(\mathbf{y}|\mathbf{X}_{j+1}
\bolds{\beta}_{j+1},\sigma _{j}^{2}\mathbf{I}_{n}\bigr)\\
&&\hspace*{17pt}\qquad{}\cdot\pi (\bolds{\beta }
_{j+1},\sigma _{j}|M_{j})\,d\bolds{\beta }_{j}\,d\sigma _{j}\biggr)\\
&&\quad \quad {}\Big/\biggl({\int
N_{n}\bigl(\mathbf{y}|\beta_{0},\sigma _{0}^{2}\mathbf{I}_{n}\bigr)\pi
(\beta _{0},\sigma _{0}|M_{0})\,d\alpha _{0}\,d\sigma _{0}}\biggr).
\end{eqnarray*}
The advantage of the posterior model probability in expression (\ref{equ1}) is that
it only involves Bayes factors for nested models. The variable selection
procedure that uses this posterior model probability as model selector is
called encompassing from below variable selection \citep{Giretal06}.
We may also use the encompassing from above approach in which all the Bayes
factors considered are of the form $B_{jk}(\mathbf{y},\mathbf{X})$ (\cite{CasMor06}). Both methods give similar results, and in this paper we
will consider the encompassing from below approach.

\begin{defn*}
Posterior model consistency when sampling from
model $M_{t}$ holds if the limit in probability $[M_{t}]$ of the random variables 
$\{\Pr (M_{j}|\mathbf{y}, \mathbf{X}),\break 
M_{j}\in \mathfrak{M}\}$ is such that
\[
\lim_{n\rightarrow \infty }\Pr (M_{j}|\mathbf{y},\mathbf{X})=\cases{
1, & \mbox{if $j=t$},
\cr
0, & \mbox{if $j\neq t$,}}\quad [M_{t}].
\]
\end{defn*}

A necessary and sufficient condition to achieve posterior model consistency
when sampling from $M_{t}$ is given in the next theorem.

\begin{theorem}\label{THM1}
When sampling from $M_{t}\in \mathfrak{M}_{t}$, posterior model consistency
holds if and only if the equality
\renewcommand{\theequation}{A}
\begin{equation}
\label{equA}\hspace*{6pt}\quad \lim_{n\rightarrow \infty }\sum_{j=0}^{k}
\mathop{\sum_{M_{j}\in \mathfrak{%
M}_{j}}}_{M_{j}\neq M_{t}}\frac{B_{j0}(\mathbf{y},\mathbf{X})}{B_{t0}(%
\mathbf{y},\mathbf{X})}
\frac{\pi (M_{j})}{\pi (M_{t})}=0,\quad [M_{t}],\hspace*{-6pt}
\end{equation}
holds.
\end{theorem}

\begin{pf}
The assertion follows from expression (\ref{equ1}).
\end{pf}

Theorem~\ref{THM1} implies that, if the Bayes factor $B_{t0}(\mathbf{y},\mathbf{X})$
is inconsistent under $M_{t}$, then posterior model consistency under $M_{t}$
does not hold. We note that when $k$ is bounded, posterior model consistency
holds for virtually any prior over the models (\cite{Casetal09}).
However, when $k=O(n^{b})$, $0<b\leq 1$, it is apparent from (\ref{equA}) that
posterior model consistency crucially depends on the rate of convergence
under $M_{t}$ to zero of the ratio $[B_{j0}(\mathbf{y},\mathbf{X})\pi
(M_{j})]/[B_{t0}(\mathbf{y},\mathbf{X})\pi (M_{t})]$.

Under the null model $M_{0}$, the necessary and sufficient condition (\ref{equA})
reduces to
%
\renewcommand{\theequation}{B}
\begin{equation}
\label{equB}\quad\lim_{n\rightarrow \infty }\sum_{j=1}^{k}
\sum_{M_{j}\in \mathfrak{M}
_{j}}B_{j0}(\mathbf{y},\mathbf{X})
\frac{\pi (M_{j})}{\pi (M_{0})}%
=0, \quad [M_{0}],\hspace*{-8pt}
\end{equation}
and it follows that if for some $M_{i}$ the Bayes factor $B_{i0}(\mathbf{y},
\mathbf{X})$ is inconsistent under $M_{0}$, then posterior model consistency
under $M_{0}$ does not hold. It is clear that, when $k=O(n^{b})$, $0<b\leq 1$,
the rate of convergence to zero of $B_{j0}(\mathbf{y},\mathbf{X})\pi (M_{j})$
determines the posterior model consistency.

Thus, from Theorem~\ref{THM1} it is clear that when $k$ increases as the sample size $%
n$ increases, posterior model consistency is a more stringent notion than
that of the Bayes factor consistency. Furthermore, posterior model
consistency depends on the specific Bayes factors $B_{j0}$ and the prior on
the class of models $\mathfrak{M}$, and, consequently, it has to be
established in a case-by-case basis.

\section{Priors and Bayes Factors for Variable~Selection}\label{sec3}

In this section we present priors for the parameters of the models and
priors over the class of models which are commonly used in variable
selection. We give formulae for the Bayes factors and their asymptotic
approximations when sampling from an arbitrary but fixed model $M_{t}$ and
rate of growth $k=O(n^{b})$ for $0\leq b\leq 1$.

\subsection{Intrinsic Priors for Model Parameters}\label{sec31}

The intrinsic priors were introduced to justify the intrinsic Bayes factor
\citep{BerPer96}. The original conditions defining the intrinsic
priors given by \citet{BerPer96} render a class of intrinsic
priors \citep{Mor97}, and a limiting procedure for choosing a specific pair
of intrinsic priors for model selection was proposed in
\citet{MorBerRac98}. This procedure is based on the additional requirement that the
intrinsic priors derived from improper priors, which are not necessarily
proper, are a limit of proper intrinsic priors.

Bayes factors for intrinsic priors were used for variable selection in
regression in \citet{MorGir05},
\citet{CasMor06},
\citet{Giretal06},
\citet{LeoMorCas12},
\citet{ConForLaR14}, among others, and this variable selection
procedure improves upon the Schwarz approximation for finite sample sizes
\citep{Casetal09} and asymptotically for high-dimensional regression
models \citep{MorGirCas10}.

The standard intrinsic method for comparing the null model $M_{0}$
{versus} the alternative $M_{j}$, starting from the improper
reference prior for the parameters of the models $M_{0}$ and $M_{j}$,
provides the proper intrinsic prior for the parameters $(\bolds{\beta}_{j+1},\sigma _{j})$, conditional on a null point $(\alpha
_{0},\sigma _{0})$, as
\begin{eqnarray*}
&& \pi^{I}(\bolds{\beta }_{j+1},\sigma _{j}|\alpha
_{0},\sigma _{0})\\
&& \quad =N_{j+1}\bigl(\bolds{
\beta}_{j+1}|\tilde{\bolds{\alpha}}_{0},\bigl(\sigma
_{j}^{2}+\sigma _{0}^{2}\bigr)
\mathbf{W}_{j+1}^{-1}\bigr)\operatorname{HC}^{+}(\sigma
_{j}|\sigma _{0}),
\end{eqnarray*}
where $\tilde{\bolds{\alpha}}_{0}=(\alpha _{0},\mathbf{0}_{j}^{\prime
})^{\prime }$, $\mathbf{W}_{j+1}^{-1}=\frac{n}{j+2}(\mathbf{X}_{j+1}^{\prime
}\mathbf{X}_{j+1})^{-1}$, and
\[
\operatorname{HC}^{+}(\sigma _{j}|\sigma _{0})=
\frac{2}{\pi }\frac{\sigma _{0}}{\sigma
_{j}^{2}+\sigma _{0}^{2}}
\]
is the half Cauchy distribution on $R^{+}$ with location parameter $0$ and
scale $\sigma _{0}$. The unconditional intrinsic prior with respect to the
reference prior $\pi ^{N}(\alpha _{0},\sigma _{0})=c_{0}/\sigma _{0}$ is
then given by
\begin{eqnarray*}
&& \pi ^{I}(\bolds{\beta}_{j+1},\sigma _{j})\\
&& \quad=\int \pi
^{I}(\bolds{\beta}_{j+1},\sigma _{j}|\alpha
_{0},\sigma _{0}) \pi^{N}(\alpha
_{0},\sigma _{0}) \,d\alpha _{0}\,d
\sigma_{0}.
\end{eqnarray*}
For comparing model $M_{0}$ {versus} $M_{j}$ the intrinsic
priors are the pair $(\pi ^{N}(\alpha _{0},\sigma _{0}),\pi ^{I}(\bolds{\beta }_{j+1},\sigma _{j}))$. We note that $\pi ^{I}(\bolds{\beta}_{j+1},\sigma _{j})$
depends\vspace*{1pt} on the arbitrary constant $c_{0}$ that
cancels out in the Bayes factor $B_{j0}(\mathbf{y},\mathbf{X})$, and hence
no tuning hyperparameters have to be adjusted. Thus, the Bayes factor for
intrinsic priors are automatically constructed from the sampling models and
the reference priors.

\subsection{Zellner's $g$-Priors for Model Parameters}\label{sec32}

For variable selection with the $g$-priors we also use the encompassing from
below approach (the encompassing from above version is given in \cite*{ScoBer10}). A basic assumption on the regression models for constructing
the $g$-priors is that the intercept and the variance error are common
parameters to all models, which reduces the number of parameters involved
when comparing $M_{j}$ {versus} $M_{0}$ from $j+4$ to $j+2$.
According to this restriction, the regression parameters of a generic model $%
M_{j}$ will be denoted as $(\beta _{0},\bolds{\beta }_{j})^{\prime }=(\beta
_{0},\beta _{1},\ldots,\beta _{j})^{\prime }$ and the variance error as~$\sigma
^{2}$, where $\beta _{0}$ and $\sigma $ are common to all models.

For a sample $(\mathbf{y},\mathbf{X})$, most references to $g$-priors in the
variable selection literature
(\cite*{BerPer01}; \cite*{ClyGeo00}; \cite*{GeoFos00}; \cite*{FerLeySte01}; \cite*{HanYu01};
\cite*{Liaetal08}, among others) refer to them as the pair $(\pi
^{N}(\beta _{0},\sigma )$, $\pi^{g}(\bolds{\beta }_{j}|\sigma))$,
where
\[
\pi ^{N}(\beta _{0},\sigma )=\frac{c_{0}}{\sigma }1_{R\times R^{+}}(
\beta _{0},\sigma )
\]
is the reference prior, and
\[
\pi (\bolds{\beta }_{j}|\sigma ,g)= N_{j}\bigl(\bolds{
\beta }_{j}| \mathbf{0}_{j},g\sigma ^{2}\bigl(
\mathbf{X}_{j}^{\prime }\mathbf{X}_{j}
\bigr)^{-1}\bigr),
\]
$g$ being an unknown positive hyperparameter, and $\mathbf{X}_{j}$ the
matrix of dimensions $n\times j$ resulting from suppressing the first column
in the design matrix $\mathbf{X}_{j+1}$ of the original formulation of the
regression model $M_{j}$.

The conjugate property of these priors makes the expression of the Bayes
factor quite simple, and it is well known that the hyperparameter $g$ plays
an important role in the behavior of the Bayes factor. Several values for $g$
have been suggested, although none of them satisfies all the reasonable
requirements (\cite*{BerPer01}; \cite*{ClyGeo04}; \cite*{ClyParVid98};
\cite*{FerLeySte01}; \cite*{GeoFos00}; \cite*{HanYu01}; \cite*{Liaetal08}). For instance, large $g$
values induce the Lindley--Bartlett paradox \citep{Bar57}, and a fixed
value for $g$ induces inconsistency, which can be corrected if $g$ were
dependent on $n$.

We consider two versions of the $g$-prior. The first is the one obtained for
$g=n$, which is justified on the ground that it provides a consistent Bayes
factor, and it is a ``unit information
prior'' \citep{KasWas95}. The second $g$-prior
version was derived for avoiding an incoherent property of the $g$-prior
detected by \citet{BerPer01}: the Bayes factor for comparing $M_{j}$
{versus} $M_{0}$ for the $g$-prior does not tend to infinity
as the coefficient of determination of $M_{j}$ tends to one. A way to avoid
this behavior is to integrate the conditional $g$-priors 
$\{\pi (\bolds{\beta }_{j}|\sigma ,g), g>0\}$ to obtain the mixture of\break $g$-priors
\[
\pi ^{\mathrm{Mix}}(\bolds{\beta }_{j}|\sigma )=\int
_{0}^{\infty }\pi (\bolds{\beta }_{j}|\sigma
,g) \pi (g)\,dg,
\]
where
\[
\pi (g)=\frac{(n/2)^{1/2}}{\Gamma (1/2)}g^{-3/2}\exp \biggl( -\frac{n}{2g}
\biggr).
\]
This mixture has been considered by some other authors, including \citet{ClyGeo04}, \citet{Liaetal08} and \citet{ScoBer10}.

\subsection{Priors for Models}\label{sec33}

Since $\mathfrak{M}$ is a discrete space, a natural default prior over it is
the uniform prior, but, as we will see, it is not a good prior when $%
k=O(n^{b})$, $1/2\leq b\leq 1$. A generalization of the uniform prior is
the parametric independent Bernoulli prior class (\cite{GeoMcC93}), for which the probability of a generic model $M_{j}$ containing $j$
out of $k$ regressors, $j\leq k$, is given by
\[
\pi (M_{j}|\theta )=\theta ^{j}(1-\theta )^{k-j},\quad
0\leq \theta \leq 1,
\]
where $\theta $ is an unknown hyperparameter, the meaning of which is the
probability of inclusion of a regressor in the model. The prior $\pi
(M_{j}|\theta )$ assigns the same probability to models with the same
dimension, and, in particular, for $\theta =1/2$ the uniform prior is obtained.

If we assume a uniform distribution for $\theta $, the unconditional
probability of model $M_{j}$ is given by
\[
\pi ^{\mathrm{HU}}(M_{j})=\int_{0}^{1}
\theta ^{j}(1-\theta )^{k-j}\,d\theta =\pmatrix{k
\cr
j}^{-1}\frac{1}{k+1}.
\]
If we decompose this probability as
\[
\pi ^{\mathrm{HU}}(M_{j})=\pi ^{\mathrm{HU}}(M_{j}|
\mathfrak{M}_{j})\pi ^{\mathrm{HU}}(\mathfrak{M}_{j}),
\]
it follows that the model prior distribution, conditional on the class $%
\mathfrak{M}_{j}$, is uniform, and the marginal over the classes $\{%
\mathfrak{M}_{j}$, $j=0,1,\ldots,k\}$ is also uniform. Then, it seems
appropriate to call to this prior the hierarchical uniform prior.

We will see that the variable selection procedure that uses the hierarchical
prior $\pi ^{\mathrm{HU}}(M_{j})$ outperforms the behavior of the one using the prior
$\pi (M_{j}|\theta )$, for any value of $\theta $.

\subsection{Bayes Factors}\label{sec34}

For the data $(\mathbf{y},\mathbf{X})$, it can easily be seen that the Bayes
factor for comparing $M_{j}$ {versus} $M_{0}$ for the $g$-prior with $g=n$ is given by
\renewcommand{\theequation}{\arabic{equation}}
\setcounter{equation}{1}
\begin{equation}
\label{equ2}
B_{j0}^{g=n}(\mathbf{y},\mathbf{X})=
\frac{(1+n)^{(n-j-1)/2}}{(1+n\mathcal{B}
_{j0})^{(n-1)/2}},
\end{equation}
for the mixture of \thinspace $g$-priors by
%
\begin{eqnarray}
&& B_{j0}^{\mathrm{Mix}}(\mathbf{y},\mathbf{X})\nonumber\\
&&\label{equ3} \quad=
\frac{(n/2)^{1/2}}{\Gamma (1/2)}
\\
&&\qquad{}\cdot\int_{0}^{\infty }
\frac{(1+g)^{(n-j-1)/2}}{(1+g\mathcal{B}_{j0})^{(n-1)/2}}%
 g^{-3/2}\exp \biggl( -\frac{n}{2g}
\biggr) \,dg,\nonumber\hspace*{-6pt}
\end{eqnarray}
and for the intrinsic priors by
%
\begin{eqnarray}
&& B_{j0}^{\mathrm{IP}}(\mathbf{y},\mathbf{X})\nonumber\\
&&\label{equ4}\quad=
\frac{2}{\pi }(j+2)^{j/2}\\
&& \qquad{}\cdot\int_{0}^{\pi /2}
\frac{\sin ^{j}\varphi (n+(j+2)\sin ^{2}\varphi )^{(n-j-1)/2}}{(n\mathcal{B}_{j0}+(j+2)\sin ^{2}\varphi)^{(n-1)/2}}
\,d\varphi.\nonumber\hspace*{-5pt}
\end{eqnarray}
The integrals on $(0,\infty )$ in (\ref{equ3}) and on $(0,\pi /2)$ in (\ref{equ4}) do not
have explicit expressions but need numerical integration.

We note that all these Bayes factors depend on the data through the
statistic $\mathcal{B}_{j0}$, which is the ratio of the square sum of the
residuals of models $M_{j}$ and $M_{0}$, that is,
%
\begin{equation}
\label{equ5}
\mathcal{B}_{j0}=\frac{\mathbf{y}^{\prime }(\mathbf{I}-\mathbf{H}_{j})%
\mathbf{y}}{\mathbf{y}^{\prime }(\mathbf{I}-({1}/{n})\mathbf{1}_{n}\mathbf{1}_{n}^{\prime })\mathbf{y}},
\end{equation}%
where $\mathbf{H}_{j}$ is the hat matrix associated to $\mathbf{X}_{j}$.

We observe that each Bayes factor exhibits a different dimension correction,
and this suggests that for small or moderate samples sizes their behavior
might be different, a point that we later explore in Section~\ref{sec5}.

For large sample sizes $n$ useful analytic approximations to the above Bayes
factors are given in the next lemma.

\begin{lemma}\label{LEM1}
For large sample sizes $n$, $k=O(n^{b})$ and $0\leq b\leq 1$, the following
approximations hold for any $j\leq k$:

\begin{longlist}[(iii)]
%
\item[(i)]\begin{equation}
\label{equ6}
\quad B_{j0}^{g=n}\approx \cases{\ds n^{-j/2}\mathcal{B}_{j0}^{-n/2}\exp \biggl\{
\frac{1}{2} \biggl( 1-\frac{1}{%
\mathcal{B}_{j0}} \biggr) \biggr\},\vspace*{1pt} \cr
\quad \mbox{for $b<1$},\vspace*{2pt}
\cr
\ds n^{-j/2}\mathcal{B}_{j0}^{-n/2}\exp \biggl\{
\frac{1}{2} \biggl( 1-\frac{1}{%
\mathcal{B}_{j0}}-\frac{j}{n} \biggr) \biggr\},\vspace*{1pt} \cr
\quad\mbox{for $b=1$},}\hspace*{-9pt}
\end{equation}

\item[(ii)]
%
\begin{equation}
\label{equ7}
\hspace*{3pt}\quad B_{j0}^{\mathrm{Mix}}\approx \cases{
\ds \biggl( \frac{n}{2} \biggr) ^{-j/2}\mathcal{B}_{j0}^{-(n-j-2)/2}
\frac{\Gamma
 (({j+1})/{2}) }{\Gamma  ({1}/{2})}, \vspace*{2pt}\cr
 \quad \mbox{for $b<1$},
\vspace*{2pt}\cr
\ds\biggl( \frac{n}{2} \biggr) ^{-j/2}\mathcal{B}_{j0}^{-(n-j-2)/2}
\vspace*{2pt}\cr
\quad\ds{}\cdot\biggl( 1+{\frac{j}{n}\mathcal{B}_{j0}} \biggr)^{-(j+1)/2}\vspace*{2pt} \cr
\quad \ds{}\cdot \frac{\Gamma
 ( ({j+1})/{2})}{\Gamma  ({1}/{2})}, \vspace*{2pt}\cr
 \quad \mbox{for $b=1$},}
\end{equation}

\item[(iii)]
%
\begin{equation}
\label{equ8}
B_{j0}^{\mathrm{IP}}\approx \cases{
\ds\biggl( \frac{n}{j+2} \biggr) ^{-j/2}\mathcal{B}_{j0}^{-(n-1)/2}
\vspace*{2pt}\cr
\ds\quad{}\cdot\exp \biggl\{ \frac{j+2}{2} \biggl( 1-\frac{1}{\mathcal{B}_{j0}} \biggr) \biggr\},
\vspace*{2pt}\cr
\quad\mbox{for $b<1$},\vspace*{2pt}\cr
\ds\biggl( 1+\frac{n}{j+2} \biggr) ^{(n-j-1)/2} \vspace*{2pt}\cr
\ds \quad{}\cdot\biggl( 1+
\frac{n\mathcal{B}_{j0}}{j+2} \biggr) ^{-(n-1)/2},\vspace*{2pt}\cr
\quad \mbox{for $b=1$}.}
\end{equation}
\end{longlist}
\end{lemma}

\begin{pf}
See Appendix~\ref{appA}.
\end{pf}

The next theorem summarizes the fact that the three Bayes factors have an
equivalent expression for large samples sizes $n$ and a bounded potential
number of regressors $k$. Further, this expression is the one obtained by
\citet{Sch78}.

\begin{theorem}\label{THM2}
When $k$ is bounded, then, for large sample sizes $n$, the Bayes factors in
(\ref{equ2}), (\ref{equ3}) and (\ref{equ4}) are equivalent to the Schwarz approximation, that is,
\[
B_{j0}^{g=n}\approx B_{j0}^{\mathrm{Mix}}\approx
B_{j0}^{\mathrm{IP}}\approx \exp \biggl( -
\frac{j}{2}\log n-\frac{n}{2}\log \mathcal{B}_{j0} \biggr).
\]
\end{theorem}

\begin{pf}
The proof follows from Lemma~\ref{LEM1} and some algebraic manipulations.
\end{pf}

Theorem~\ref{THM2} implies that for low-dimensional regular models, any Bayes factor
is consistent, as the Schwarz approximation guarantees Bayes factor
consistency. In this setting, for any positive model prior, posterior model
consistency under an arbitrary model $M_{t}$ also holds.

However, for high-dimensional models the Schwarz approximation does not
necessarily guarantee either the Bayes factor consistency\ or the posterior
model consistency. Other approximating forms than that of Schwarz appear
in this latter setting.

\subsection{Asymptotic Approximations to the Bayes~Factors}\label{sec35}

The Bayes factor approximations in (\ref{equ6}), (\ref{equ7}) and (\ref{equ8}) depend on the random
sequence $\{\mathcal{B}_{j0},n\geq 1\}$ given in~(\ref{equ5}). In this section we go
a step forward and use the asymptotic distribution of the statistic $\mathcal{B}_{j0}$ under an arbitrary but fixed model $M_{t}$ to approximate
the Bayes factors $B_{j0}^{g=n}$, $B_{j0}^{\mathrm{Mix}}$ and $B_{j0}^{\mathrm{IP}}$.

We first note that the asymptotic distribution of $\mathcal{B}_{j0}$ under $%
M_{t}$, a doubly noncentral beta distribution, depends on the limit of the
pseudo-distance between models defined as
\[
\delta _{n}(M_{t},M_{j})=\frac{1}{2 \sigma _{t}^{2}}
\bolds{\beta}_{t}^{\prime}\frac{\mathbf{X}_{t}^{\prime }(\mathbf{I}_{n}-\mathbf{H}
_{j})\mathbf{X}_{t}}{n}\bolds{\beta}_{t}.
\]
General properties of this pseudo-distance have been studied in
\citet{Giretal10}. This pseudo-distance $\delta _{n}(M_{t},M_{j})$ can be
simplified as follows. We first write the covariance matrix of the joint set
of covariates of the model $M_{t}$ and $M_{j}$, the dimensions of which are
$(t+j)\times (t+j)$, as
\[
\Sigma _{t+j}^{(n)}=
\pmatrix{ S_{tt}^{(n)}
& S_{tj}^{(n)}
\vspace*{2pt}\cr
S_{jt}^{(n)} &
S_{jj}^{(n)}},
\]
where the matrices $S_{tt}^{(n)}$, $S_{jj}^{(n)}$ are definite positive.
Let us consider the matrices $S_{t\cdot j}^{(n)}=S_{tt}^{(n)}-S_{tj}^{(n)}{S_{jj}^{(n)}}^{-1}S_{jt}^{(n)}$, and $S_{t \cdot j}=\lim_{n\rightarrow \infty
}S_{t \cdot j}^{(n)}$. Then, it can now be seen that $\delta _{n}(M_{t},M_{j})$
can be expressed as
\[
\delta _{n}(M_{t},M_{j})=\frac{1}{2\sigma _{t}^{2}}
\bolds{\beta}_{t}^{\prime }S_{t \cdot j}^{(n)}
\bolds{\beta }_{t}.
\]

In what follows we denote $\delta ^{\ast }(M_{t},M_{j})
= \break \lim_{n\rightarrow \infty }\delta _{n}(M_{t},M_{j})$, and if there is no
confusion, $\delta^{\ast }(M_{t},M_{j})$ and $\delta _{n}(M_{t},M_{j})$ will
be simply written as $\delta _{tj}^{\ast }$ and\vspace*{1pt} $\delta _{tj}$.

For any $M_{j}$, using the asymptotic distribution of $\mathcal{B}_{j0}$
under $M_{t}$, we can now provide asymptotic approximations in probability
$[M_{t}]$ to the Bayes factors $B_{j0}^{g=n}$, $B_{j0}^{\mathrm{Mix}}$ and
$B_{j0}^{\mathrm{IP}}$ that we summarize in Lemma~\ref{LEM2}.

\begin{lemma}\label{LEM2}
When sampling from a model $M_{t}$, the Bayes factors in (\ref{equ2}), (\ref{equ3}) and (\ref{equ4})
for $j\leq k=O(n^{b})$, $b\leq 1$, can be approximated for large $n$ as
%
\begin{eqnarray}
\label{equ9}
B_{j0}^{g=n} &\approx &  \cases{\ds
n^{-j/2} \biggl( \frac{1+\delta _{tj}^{\ast }}{1+\delta _{t0}^{\ast }} \biggr) ^{-n/2}\vspace*{2pt}\cr
\ds\quad {}\cdot\exp \biggl(
\frac{\delta _{tj}^{\ast }-\delta _{t0}^{\ast }}{2(1+\delta _{tj}^{\ast })} \biggr), \hspace*{2pt}\cr
\quad\mbox{for $b<1$},
\vspace*{2pt}\cr
\ds n^{-j/2} \biggl( \frac{1+\delta _{tj}^{\ast }-j/n}{1+\delta _{t0}^{\ast }}
 \biggr) ^{-n/2}\vspace*{2pt}\cr
\ds\quad{}\cdot\exp \biggl( \frac{\delta _{tj}^{\ast }-\delta _{t0}^{\ast
}-j/n}{2(1+\delta _{tj}^{\ast }-j/n)} \biggr),\vspace*{2pt} \cr
\quad\mbox{for $b=1$},}
\\
\label{equ10}
B_{j0}^{\mathrm{Mix}} &\approx &  \cases{
\ds\biggl( \frac{n e}{j+1} \biggr) ^{-j/2} \biggl( \frac{1+\delta
_{tj}^{\ast }}{1+\delta _{t0}^{\ast }}
\biggr) ^{-(n-j-2)/2},\vspace*{2pt} \cr
\ds\quad\mbox{for $b<1$},
\vspace*{2pt}\cr
\ds \biggl( \frac{n e}{j+1} \biggr) ^{-j/2} \vspace*{2pt}\cr
\quad\ds{}\cdot\biggl( \frac{1+\delta
_{tj}^{\ast }-j/n}{1+\delta _{t0}^{\ast }}
\biggr) ^{-(n-j-2)/2},\vspace*{2pt} \cr
\quad \mbox{for $b=1$},}\hspace*{-16pt}
\end{eqnarray}
and
%
\begin{equation}
\label{equ11}
\quad B_{j0}^{\mathrm{IP}}\approx \cases{
\ds\biggl( \frac{n}{j+2} \biggr) ^{-j/2} \biggl( \frac{1+\delta _{tj}^{\ast }}{%
1+\delta _{t0}^{\ast }}
\biggr) ^{-(n-j)/2}, \vspace*{2pt}\cr
\quad\mbox{for $b<1$},
\vspace*{2pt}\cr
\ds \biggl( 1+\frac{n }{j} \biggr) ^{(n-j-1)/2}\vspace*{2pt} \cr
\ds \quad{}\cdot\biggl(
\frac{({n}/{j})%
(1+\delta _{tj}^{\ast })+\delta _{t0}^{\ast }}{1+\delta _{t0}^{\ast }}%
 \biggr) ^{-(n-1)/2}, \vspace*{2pt}\cr
 \quad\mbox{for $b=1$.}}\hspace*{-6pt}
\end{equation}
\end{lemma}

\begin{pf}
The proof follows from Lemma~\ref{LEM1} and the asymptotic distribution of the
statistic $\mathcal{B}_{j0}$ under model~$M_{t}$ \citep{Casetal09}, and
it is omitted.
\end{pf}

From Lemma~\ref{LEM2} it follows that when sampling from the null model $M_{0}$, that
is, when $M_{t}=M_{0}$, the asymptotic approximations (\ref{equ9}), (\ref{equ10}) and (\ref{equ11})
notably simplify, as they only depend on $n$ and the dimension $j$ of the
model, irrespective of the particular set of covariates. This means that,
under the null model $M_{0}$, the above Bayes factors are asymptotically
constant across models in the class $\mathfrak{M}_{j}$.

To prove some results on posterior model consistency when sampling from an
alternative model $M_{t}$, we need to know for which models $M_{j}$ in
$\mathfrak{M}$ the pseudo-distance $\delta^{\ast}(M_{t},M_{j})$ is zero.
This result follows from Lemma~\ref{LEM3}.

\begin{lemma}\label{LEM3}
\textup{(i)} For any model $M_{j}$ such that $\dim (M_{j})<$ $\dim (M_{t})$, we have
that
\[
\delta^{\ast }(M_{t},M_{j})>0.
\]

\textup{(ii)} For any model $M_{j}$ such that $\dim (M_{j})= \dim (M_{t})$,
\[
\delta ^{\ast }(M_{t},M_{j})=\cases{0, &
\mbox{if $M_{j}=M_{t}$,}
\vspace*{2pt}\cr
{>}0, & \mbox{if $M_{j}\neq M_{t}$.}}
\]

\textup{(iii)} For any model $M_{j}$ such that $\dim (M_{j})> \dim (M_{t})$,
\[
\delta ^{\ast }(M_{t},M_{j})= \cases{0, & \mbox{if $M_{t}$ is nested in $M_{j}$},
\cr
{>}0,  & \mbox{otherwise}.}
\]
\end{lemma}

\begin{pf}
We note that (a) if $M_{t}$ and $M_{j}$ do not have common covariates, then
the matrix $\Sigma _{t+j}=\lim_{n\rightarrow \infty }\Sigma _{t+j}^{(n)}$ is\vspace*{1pt}
positive definite, and hence $S_{t \cdot j}$ is positive definite, and (b) if
$M_{t}$ and $M_{j}$ do have common covariates, then it can be seen that
\[
S_{t \cdot j}= \pmatrix{\mathbf{P} & \mathbf{O}
\cr
\mathbf{O} & \mathbf{O}},
\]
where $\mathbf{P}$ is a positive definite matrix of dimensions $\max
\{0,\dim M_{t}-\dim M_{j}\}$. We observe that if either $\dim M_{t}=\dim
M_{j}$ or $M_{t}$ is nested in $M_{j}$, we have that $S_{t \cdot j}=\mathbf{O}$.
The proof of Lemma~\ref{LEM3} follows from (a) and~(b) and the fact that all regression
coefficients $\bolds{\beta}_{t}$ in model $M_{t}$ are different from
zero.
\end{pf}

It is interesting to remark that for $b<1$ and any $M_{j}$ such that
$\delta^{\ast }(M_{t},M_{j})>0$, the rate of convergence in probability $[M_{t}]$
to zero\vspace*{1pt} of $B_{j0}^{g=n}$, $B_{j0}^{\mathrm{Mix}}$\vspace*{1pt} and $B_{j0}^{\mathrm{IP}}$ for $M_{t}\neq
M_{0}$ is exponentially fast, but the rate of convergence in probability $%
[M_{0}]$ to zero for $j\neq 0$ is only potentially fast. This is in
line with the result for $b=0$ obtained by \citet{Daw11} (for discrete data
see \cite*{ConForLaR14}).

\section{Posterior Model Consistency for $k=O(n^{b})$ and \texorpdfstring{$0\leq b\leq
1$}{0 leq b leq 1}}\label{sec4}

Posterior model consistency results for the six Bayesian variable selection
procedures defined by the Bayes factors $B_{j0}^{g=n}$, $B_{j0}^{\mathrm{Mix}}$,
$B_{j0}^{\mathrm{IP}}$, the Bernoulli model prior $\pi (M_{j}|\theta )$ and the
hierarchical uniform prior $\pi ^{\mathrm{HU}}(M_{j})$, when sampling from an
arbitrary but fixed model $M_{t}$ are summarized in Theorem~\ref{THM3}. For
simplicity, the posterior model consistency results for the case when sampling
from model $M_{0}$ and from an alternative model $M_{t}$ are not separated.
However, we keep in mind that the rate of convergence of the posterior model
probabilities when sampling from $M_{0}$ is different from the rate of
convergence when sampling from $M_{t}\neq M_{0}$.

\begin{theorem}\label{THM3}
\textup{(i)} When sampling from $M_{t}$ and $k=O(n^{b})$, the Bayesian\vspace*{1pt} procedures for
the Bayes factors $B_{j0}^{g=n}$, $B_{j0}^{\mathrm{Mix}}$, $B_{j0}^{\mathrm{IP}}$, and the
Bernoulli\vspace*{1pt} model prior are posterior model consistent for $0\leq b<1/2$ and
posterior model inconsistent for $1/2\leq b\leq 1$.

\textup{(ii)} When sampling from $M_{t}$ and $k=O(n^{b})$, the Bayesian procedures
for the Bayes factors $B_{j0}^{g=n}$, $B_{j0}^{\mathrm{Mix}}$ and $B_{j0}^{\mathrm{IP}}$ and
the hierarchical uniform prior are posterior model consistent for $0\leq
b\leq 1$.
\end{theorem}

\begin{pf}
See Appendix~\ref{appB}.
\end{pf}

It is interesting to observe that the Bernoulli prior $\pi (M_{j}|\theta )$,
conditional on $\theta $, induces a Binomial distribution on the classes
$\mathfrak{M}_{j}$, which, in turn, by the change of variables $x=j/k$,
induces a distribution on $x\in [ 0,1]$ which converges in probability
to a Dirac's delta on $\theta $, as $k$ tends to infinity. In other words,
for large values of $k$ the Bernoulli prior concentrates around models which
have a proportion of covariates close to~$\theta$. Therefore, this
apparently innocuous prior conveys too much prior information about the
proportion of covariates of the models, and thus it makes the posterior
model probabilities for $1/2\leq b\leq 1$ inconsistent. This wrong
asymptotic behavior is corrected by the hierarchical uniform prior.

\section{Small Sample Comparisons}\label{sec5}

Given a Bayes factor $B_{j0}$ for the models $\{M_{0},M_{j}\}$, the decision
of choosing model $M_{j}$ when $\Pr (M_{j}| \break B_{j0})\geq 1/2$ is an optimal
decision under $\Pr (M_{0})=\Pr (M_{j})=1/2$ and a $0\mbox{--}1$ loss function. We
recall that for a uniform prior on the class of models $\mathfrak{M}$, to
rank the models in the class according to their posterior model
probabilities is equivalent to the ranking produced by the Bayes factor. In
spite of this, a sampling analysis of the optimal statistical decision
function has been long claimed [see, e.g., \citet{Fra11} and
discussions therein]. From expression (\ref{equ2}), (\ref{equ3}) and (\ref{equ4}) it is apparent that
the dimension correction of the Bayes factors for the $g$-prior with $g=n$,
for the mixture of $g$-priors and for the intrinsic priors are different
from each other. This suggests that their sampling behavior for small and
moderate sample sizes might be different.

\begin{figure}

\includegraphics{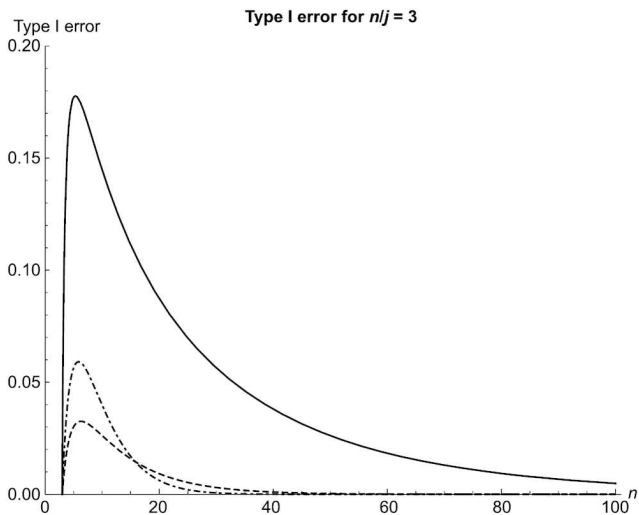}

\caption{Type I error probabilities for the intrinsic procedure (continuous line),
the $g$-prior with $g=n$ (dot-dashed) and the mixture of $g$-priors (dashed).}
\label{fig1}
\end{figure}

In this section we study the sampling properties of the posterior model
probabilities for $\Pr (M_{0})=\Pr (M_{j})=1/2$ and the Bayes factors $B_{j0}^{g=n},B_{j0}^{\mathrm{Mix}}$ and $B_{j0}^{\mathrm{IP}}$. We recall\vspace*{1pt} that the posterior
probability $\Pr (M_{j}|\mathbf{y,X})$ for any of these Bayes factors
depends on the data $(\mathbf{y,X})$ through the same statistic $\mathcal{B}%
_{j0}$, which takes values in the interval $(0,1)$. Therefore, the critical
regions for rejecting the null model $M_{0 }$ for these Bayesian
procedures are
\begin{eqnarray*}
R_{j0}^{(g=n)} &=& \bigl\{\mathcal{B}_{j0}\dvtx \Pr
\bigl(M_{j}|B_{j0}^{g=n}\bigr)\geq 1/2\bigr\},
\\
R_{j0}^{\mathrm{Mix}} & =& \bigl\{\mathcal{B}_{j0} \dvtx  \Pr
\bigl(M_{j}|B_{j0}^{\mathrm{Mix}}\bigr)\geq 1/2\bigr\}
\end{eqnarray*}
and
\[
R_{j0}^{\mathrm{IP}}=\bigl\{\mathcal{B}_{j0} \dvtx \Pr
\bigl(M_{j}|B_{j0}^{\mathrm{IP}}\bigr)\geq 1/2\bigr\}.
\]%
These critical regions are in $(0,1)$ and, since the posterior probabilities
are monotone increasing functions of $\mathcal{B}_{j0}$, they are intervals.
Using the distribution of the statistic $\mathcal{B}_{j0}$ under $M_{0}$ and
$M_{j}$, we can compute the exact value of the Type I and II errors
probabilities as a function of the model dimension $j$ and the sample size $n$.
Figure~\ref{fig1} shows the Type I error probabilities of the optimal decision
rule associated to the regions $R_{j0}^{(g=n)},R_{j0}^{\mathrm{Mix}}$ and $%
R_{j0}^{\mathrm{IP}} $ for $j=n/3$\thinspace and the sample size $n=1,\ldots,100$.

From Figure~\ref{fig1} it follows that the Type I error probabilities of the
procedures based on $g$-priors are very close to each other and smaller
than that based on the intrinsic priors. We note that as $n$ and $j$
increase at the same pace, $n/j=3$, the Type I error probabilities for the
procedure based on $g$-priors go faster to zero than the procedure based on
the intrinsic priors does.

\begin{figure}

\includegraphics{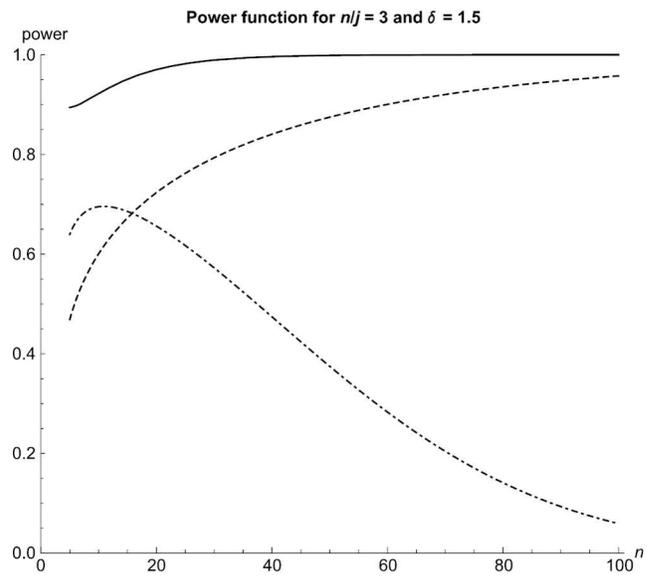}

\caption{Power for the Bayes factor for intrinsic priors (continuous line), for
\thinspace $g$-priors for $g=n$ (dot-dashed) and for the mixture of $g$-priors (dashed).}\label{fig2}
\end{figure}

In Figure~\ref{fig2} we display for $\delta _{j0}=1$ and $j=n/3$ the power function
of the above procedures as a function of the sample size $n=1,\ldots,100$.

From Figure~\ref{fig2} we observe that the power of the procedure based on intrinsic
priors is much larger than those based on the $g$-priors. This is the price
the procedures based on $g$-priors pay for their very small Type I error
probabilities. Further, the power for the intrinsic priors and the mixture
of $g$-priors increases to one as the sample size $n$ and the model dimension
$j$ increase at the same pace, that is, $n/j=r\geq 1$, but the power for the
$g$-prior with $g=n$ increases as $n$ increases up to a certain $n$ and then
decreases, which is a surprisingly unreasonable behavior. The explanation to
the anomalous behavior of the Bayes factor for the $g=n$ is due to the
inconsistency of this Bayes factor for any model $M_{j}$ such that $j=O(n)$,
a point that we discuss in Section~\ref{sec6} and summarize in Table~\ref{tab2}.

On the other hand, we remark that as $\delta _{j0}$ increases, the power of
the three procedures increases for any sample size $n$.

Figures~\ref{fig1} and~\ref{fig2} indicate how unbalanced are the Type I and II error
probabilities of the Bayesian procedures based on $g$-priors compared with
that based on the intrinsic priors. The practical implications of this
analysis are that for moderate sample sizes the Bayesian procedures based on $%
g$-priors are strongly biased toward the null model.

\section{Concluding Remarks}\label{sec6}

Variable selection in regression is a central problem in statistical
inference and the aim of this paper has been to evaluate the sampling
properties of Bayesian model selection procedures, a requirement long
advocated by many statisticians. For some interesting applications the
number of regressors is very large, and hence we assumed that the potential
number of regressors $k$ grows with $n$. We very soon realized that the
variable selection takes place in a large class of models, and hence
posterior model consistency seems to be the appropriate asymptotic property
to be explored, a~concept that depends on the priors over the models and the
model parameters. Posterior model consistency for variable selection for
three popular Bayes factors and two types of model priors has been explored,
although the methodology we used can be applied to any other specific Bayes
factor and model prior.

For low-dimensional normal regression models it is well known that virtually
any Bayes factor has an asymptotic approximation which is equivalent to the
Schwarz approximation, which assures consistency. However, for large-dimensional models more appropriate asymptotic approximations for the Bayes
factors, such as those given in Lemma~\ref{LEM2}, are necessary for analyzing consistency.
\begin{table}[b]
\caption{Posterior model consistency
when sampling from $M_{t}$,
as a~function of the Bayes factor, model\hspace*{1pt} prior and the rate  of growth of
$k=O ({n}^{b}\bigr)$}\label{tab1}
\begin{tabular*}{\columnwidth}{@{\extracolsep{\fill}}lcc@{}}
\hline
 \textbf{Model} \textbf{prior}
&
$\bolds{\pi(M|\theta )}$ & $\bolds{\pi^{\mathrm{HU}}(M)}$
\\
\hline
Bayes factor
 & $B_{j0}^{g=n}$, $B_{j0}^{\mathrm{Mix}}$, $B_{j0}^{\mathrm{IP}}$ & $B_{j0}^{g=n}$,
$B_{j0}^{\mathrm{Mix}}$, $B_{j0}^{\mathrm{IP}}$
\\
 $0\leq b< 1/2$
 & \multicolumn{2}{c@{}}{Consistent}
\\
$1/2\leq b\leq 1$
 & Inconsistent & Consistent \\
\hline
\end{tabular*}
\end{table}

Although we considered the independent Bernoulli class of model\vspace*{1pt} priors
$\{\pi (M|\theta )$, $\theta \in (0,1),M\in \mathfrak{M}\}$ and the
hierarchical uniform prior $\pi ^{\mathrm{HU}}{\small (M)}$, a mixture of $\pi
(M|\theta )$ with respect to the uniform distribution on~$\theta $, the
asymptotic results for the hierarchical uniform prior can be formally
extended to other regular mixtures of Bernoulli model priors.

The conclusions on posterior model consistency we draw for the above
Bayesian procedures when sampling from an arbitrary but fixed model $M_{t}$
and for different rates of growth of $k$ are summarized in Table~\ref{tab1}.

Table~\ref{tab1} implies that when sampling from $M_{t}$, the Bayesian procedures for
the Bayes factors $B_{j0}^{g=n}$, $B_{t0}^{\mathrm{Mix}}$ and $B_{t0}^{\mathrm{IP}}$ and the
Bernoulli model prior are inconsistent for $1/2\leq b\leq 1$, but for the
hierarchical uniform model prior they are consistent for any $0\leq b\leq 1$.
Thus, a first conclusion is that the hierarchical uniform model prior $\pi
^{\mathrm{HU}}(M)$ outperforms the independent Bernoulli model prior $\pi (M|\theta )$.

We remark that the above results are valid when sampling from a fixed model $M_{t}$ with finite dimension. The analysis of the infinite dimensional case
is an open problem that deserves more efforts, as the Bayes factors are now
not necessarily consistent \citep{MorGirCas10} and, consequently, the
posterior model consistency results differ from those presented above. For
instance, for $t=O(n)$, the Bayes factor $B_{t0}^{g=n}$ is such that
\[
\lim_{n\rightarrow \infty }B_{t0}^{g=n}=0,\quad [M_{t}],
\]%
so that it is inconsistent under any model $M_{t}\neq M_{0}$, and this
implies that it is also posterior model inconsistent under $M_{t}\neq M_{0}$
for any model prior.

For the Bayes factors $B_{t0}^{\mathrm{Mix}}$ and $B_{t0}^{\mathrm{IP}}$ the situation is not
so dramatic. The set of alternative models $M_{t}$ for which inconsistency
of $B_{t0}^{\mathrm{Mix}}$ holds is a small set of models around $M_{0}$ that satisfy
the condition
\[
\delta _{t0}^{\ast }<\delta _{\mathrm{mix}}(r)= \biggl( 1-
\frac{1}{r} \biggr) (e
r)^{1/(r-1)}-1,
\]
where $r=n/t>1$. Likewise, the set of alternative models $M_{t}$ for which $%
B_{t0}^{\mathrm{IP}}$ is inconsistent is that given by the condition
\[
\delta _{t0}^{\ast }(r)<\delta _{\mathrm{IP}}(r)=
\frac{r-1}{(r+1)^{(r-1)/r}}-1.
\]%
A summary of these results is given in Table~\ref{tab2}.
\begin{table}[b]
\caption{Posterior model
consistency when sampling from $M_{t}$,  for
$t=n/r$, $r>1$, and $\pi (M_{0})>0$}\label{tab2}
\begin{tabular*}{\columnwidth}{@{\extracolsep{\fill}}lcl@{}}
\hline
\textbf{Bayes factor} &
\textbf{Model prior} & \multicolumn{1}{c@{}}{\textbf{Posterior
model consistency}}
\\
\hline
$B_{t0}^{g=n}$ \vspace*{3pt}
& $\pi (M_{0})>0$ &  \hspace*{6pt}Inconsistent under any $M_{t}$ \\
$B_{t0}^{\mathrm{Mix}}$
& $\pi (M_{0})>0$ &
\hspace*{12pt}Inconsistent under $M_{t}$
\\
&& \hspace*{12pt}such that $\delta _{t0}^{\ast }<\delta_{\mathrm{mix}}(r)$
\\
$B_{t0}^{\mathrm{IP}}$
& $\pi (M_{0})>0$ &
\hspace*{12pt}Inconsistent under $M_{t}$
\\
&& \hspace*{12pt}such that $\delta _{t0}^{\ast }<\delta_{\mathrm{IP}}(r)$
\\
\hline
\end{tabular*}
\end{table}

From Table~\ref{tab2} we can draw the conclusion that the intrinsic priors and the
mixture of $g$-priors are preferred to the  $g$-prior for $g=n$.

We also note that $\delta _{\mathrm{IP}}(r)<\delta _{\mathrm{mix}}(r)$, so that the
inconsistency region of the Bayes factor for the intrinsic priors is smaller
than that for the mixture of $g$-priors. Further, for the case where $r=1$
it can be shown that the Bayes factor $B_{t0}^{\mathrm{Mix}}$ is inconsistent for
any alternative model $M_{t}$, while the Bayes factor $B_{t0}^{\mathrm{IP}}$ is
inconsistent only for those $M_{t}$ such that $\delta _{t0}^{\ast }<1/\log 2-1$.

On the other hand, for small and moderate sample sizes, Figures~\ref{fig1} and \ref{fig2} that we
presented indicate that the behavior of the Bayes factors $B_{t0}^{g=n}$and $%
B_{j0}^{\mathrm{Mix}}$ are strongly biased toward the null model, while the Bayes
factor for the intrinsic priors $B_{j0}^{\mathrm{IP}}$ has more balanced Type I and
II error probabilities.

Therefore, the overall conclusion from our analysis is that the intrinsic
priors over the model parameters and the hierarchical uniform prior over the
models are nowadays the priors to be recommended for variable selection in
normal regression.

\begin{appendix}


\section{Proof of Lemma~\texorpdfstring{\protect\ref{LEM1}}{1}}\label{appA}

Part (i) is immediate and hence it is omitted. Part (ii) follows by first
making the change of variables $y=\exp [-n/(2g)]$ in the integral in (\ref{equ3}).
The Jacobian of the inverse transformation is
\[
J=\frac{dg}{dy}=\frac{n}{2y\log y^{2}}
\]%
and, thus, the integral in (\ref{equ3}) now becomes
\begin{eqnarray*}
&& \int_{0}^{1} \biggl( 1-\frac{n}{2\log y}
\biggr) ^{(n-j-1)/2} \biggl( 1-\frac{n%
\mathcal{B}_{j0}}{2\log y} \biggr) ^{(-n+1)/2}
\\
&&\hspace*{4pt}\quad{}\cdot\biggl( -\frac{n}{2\log y}%
 \biggr) ^{-3/2}y J\,dy.
\end{eqnarray*}
The first factor in this integral can be approximated by
\begin{eqnarray*}
&& \biggl( 1-\frac{n}{2\log y} \biggr) ^{(n-j-1)/2}\\
&& \quad\approx y^{-1+({j}/{n})
}
\biggl( -\frac{n}{2\log y} \biggr) ^{(n-j-1)/2}
\end{eqnarray*}
and the second by
\begin{eqnarray*}
&& \biggl( 1-\frac{n\mathcal{B}_{j0}}{2\log y} \biggr) ^{(-n+1)/2}\\
&& \quad\approx
\mathcal{B}_{j0}^{(1-n)/2}y^{{1}/{\mathcal{B}_{j0}}} \biggl( -
\frac{n}{%
2\log y} \biggr) ^{(1-n)/2}.
\end{eqnarray*}
Plugging these approximations in the integral, and after some
simplifications, we obtain that the original Bayes factor can be
approximated as
\begin{eqnarray*}
&& B_{j0}^{\mathrm{Mix}}(\mathbf{y},\mathbf{X})\\
&& \quad \approx
\frac{(n/2)^{-j/2}}{\Gamma (1/2)}%
\mathcal{B}_{j0}^{(-n+1)/2}\\
&& \qquad {}\cdot\int
_{0}^{1}y^{({1}/{\mathcal{B}_{j0}})-1+({j}/{n})} \biggl( -
\frac{1}{\log y} \biggr) ^{(1-j)/2}\,dy.
\end{eqnarray*}
For any $j$ and $n$, the integral in this expression has value
\begin{eqnarray*}
&& \int_{0}^{1}y^{({1}/{\mathcal{B}_{j0}})-1+({j}/{n})} \biggl( -
\frac{1}{%
\log y} \biggr) ^{(1-j)/2}\,dy\\
&& \quad =\mathcal{B}_{j0}^{(j+1)/2}
\biggl( 1+\frac{j}{n}%
\mathcal{B}_{j0} \biggr)
^{-(j+1)/2}{\large \Gamma } \biggl( \frac{j+1}{2}%
 \biggr) ,
\end{eqnarray*}
and thus the approximation of the Bayes factor is
\begin{eqnarray*}
&& B_{j0}^{\mathrm{Mix}}(\mathbf{y},\mathbf{X})\\
&& \quad \approx \biggl(
\frac{n}{2} \biggr) ^{-j/2}%
\mathcal{B}_{j0}^{-(n-j-2)/2}
\biggl( 1+{\frac{j}{n}\mathcal{B}_{j0}} \biggr) ^{-(j+1)/2}\\
&& \qquad{}\cdot\frac{\Gamma  ( ({j+1})/{2} ) }{\Gamma  (
{1}/{2} ) }.
\end{eqnarray*}
If $b<1$, we have that
\[
\lim_{n\rightarrow \infty } \biggl( 1+\frac{j}{n}\mathcal{B}_{j0}
\biggr) ^{-(j+1)/2}=1,
\]%
and this proves the first part of (ii). If $b=1$, the proof follows suit
directly from the expression of the approximation. This completes the proof
of part (ii).

Part (iii) was proved in \citet{Giretal10} and hence it is
omitted.

\section{Proof of Theorem~\texorpdfstring{\protect\ref{THM3}}{3}}\label{appB}

\begin{longlist}[5.]
\item[1.]  We first prove that condition (\ref{equA}) holds for the Bayes factor
$B_{j0}^{g=n}$, the Bernoulli\vspace*{1pt} model prior $\pi (M_{j}|\theta )$ and $0\leq
b<12$, and that it does not hold for $1/2\leq  b\leq 1$. For, under the
Bernoulli prior we have that
\begin{eqnarray*}
\mbox{(\ref{equA})}&=&\sum_{j=0}^{k}\mathop{\sum
_{ M_{j}\in \mathfrak{M}_{j}}}_{M_{j}\neq M_{t}}n^{({t-j})/{2}} \bigl( 1+\delta
_{tj}^{\ast } \bigr) ^{-n/2}\\
&&\hspace*{47pt}{}\cdot\exp \biggl\{
\frac{1}{2} \biggl( \frac{\delta _{tj}^{\ast }-\delta _{t0}^{\ast }}{%
1+\delta _{tj}^{\ast }} \biggr) \biggr\} \biggl(
\frac{\theta }{1-\theta }%
 \biggr) ^{j-t}.
\end{eqnarray*}
From Lemma~\ref{LEM3}, the terms for $j\leq t$ go to zero as $n$ tends to infinity.
For $j>t$ let us split the class $\mathfrak{M}_{j}$ as
\[
\mathfrak{M}_{j}=\mathfrak{N}_{j}\cup (
\mathfrak{M}_{j}-\mathfrak{N}_{j}),
\]%
where $\mathfrak{N}_{j}$ is the class of models $M_{j}$ such that $M_{t}$ is
nested in $M_{j}$. From Lemma~\ref{LEM3}, it follows that $\delta _{tj}^{\ast }=0$
for any $M_{j}\in \mathfrak{N}_{j}$, and $\delta _{tj}^{\ast }>0$ for $%
M_{j}\in \mathfrak{M}_{j}-\mathfrak{N}_{j}$. Therefore, for large $n$ the
contribution of the models in $\mathfrak{M}_{j}-\mathfrak{N}_{j}$ to the sum
in (\ref{equA})\vadjust{\goodbreak} tends to zero, and we then have for large $n$ that
\begin{eqnarray*}
\mbox{(\ref{equA})} &\approx &\sum_{j=t+1}^{k}\sum
_{M_{j}\in \mathfrak{N}_{j}}n^{({t-j})/{2}}\\
&&\hspace*{32pt}\qquad{}\cdot \exp \biggl\{ \frac{1}{2} \biggl(
\frac{\delta _{tj}^{\ast }-\delta
_{t0}^{\ast }}{1+\delta _{tj}^{\ast }} \biggr) \biggr\} \biggl( \frac{\theta }{%
1-\theta } \biggr)
^{j-t}
\\
&\approx &\sum_{i=1}^{k-t}\pmatrix{k-t
\cr
i}n^{-{i}/{2}} \biggl( \frac{\theta
}{1-\theta } \biggr) ^{i}
\\
&=& \biggl( -1+ \biggl( 1+\frac{\theta }{(1-\theta )n^{1/2}} \biggr) ^{k-t} \biggr)
\\
&\approx &  \exp \bigl\{ n^{b-1/2} \bigr\}.
\end{eqnarray*}%
Then, for large $n$, (\ref{equA}) is equivalent to $\exp \{n^{b-1/2}\}$ and this
proves the assertions.

\item[2.]  We now prove that for the Bayes factor $B_{j0}^{g=n}$ and the
hierarchical uniform prior $\pi ^{\mathrm{HU}}(M)$, condition (\ref{equA}) does hold for any
$b\leq 1$. Indeed, for large $n$, using again the decomposition $\mathfrak{M}%
_{j}=\mathfrak{N}_{j}\cup (\mathfrak{M}_{j}-\mathfrak{N}_{j})$, the sum (\ref{equA})
can be approximated for large $n$ as
\begin{eqnarray*}
\mbox{(\ref{equA})} &\approx & \sum_{j=t+1}^{k}\sum
_{M_{j}\in \mathfrak{N}_{j}}n^{({t-j})/{2}
}\exp \biggl\{ \frac{1}{2} \biggl(
\frac{\delta _{tj}^{\ast }-\delta
_{t0}^{\ast }}{1+\delta _{tj}^{\ast }} \biggr) \biggr\}
\frac{{k\choose t}}{{k \choose j}}
\\
&=&\sum_{i=1}^{k-t}n^{-{i}/{2}}
\frac{(t+i)!}{i!}
\\
&<&\sum_{i=1}^{\infty }n^{-{i}/{2}}
\frac{(t+i)!}{i!}
\\
&=&t! \biggl( -1+\bigl(1-n^{-1/2}\bigr)^{-t}\frac{n^{1/2}}{n^{1/2}-1}
\biggr) .
\end{eqnarray*}
The last expression tends to zero as $n$ tends to infinity, and this proves
the assertion.

\item[3.]  Let us now consider the Bayes factor $B_{j0}^{\mathrm{IP}}$ and the Bernoulli
prior. For simplicity we prove the assertion for $\theta =1/2$, as the proof
for any $\theta $ follows the same line of reasoning. We first note that the
contribution of the models in $\mathfrak{M}_{j}-\mathfrak{N}_{j}$ to the sum
in (\ref{equA}) tends to zero as $n$ tends to infinity. Thus, we have for large $n$
that
\[
\mbox{(\ref{equA})}\approx (t+2)^{-t/2}\sum_{j=t+1}^{k}
\biggl( \frac{n}{j+2} \biggr) ^{-j/2}n^{t/2}\pmatrix{k-t
\cr
j-t},
\]%
which, after the change of variables $i=j-t$, adopts the form
\begin{eqnarray*}
\mbox{(\ref{equA})} &\approx &  (t+2)^{-t/2}\\
&& {}\cdot\sum_{i=1}^{k-t}a_{i}
\frac{(k-t)(k-t-1)\cdots (k-t-i+1)%
}{n^{i/2}},
\end{eqnarray*}
where
\[
a_{i}=\frac{(t+i+2)^{({t+i})/{2}}}{i!}.
\]%
It can be seen that the sequence $\{a_{i}\}$ increases as $i$ increases for
$i<i_{0}(t)$, where $i_{0}(t)\approx [ 1+1.65\sqrt{t}]$, and decreases
for $i>i_{0}(t)$, and thus it is bounded by some function of $t$, say, $a(t)$. Thus, the sum in (\ref{equA}) is upper bounded as
\begin{eqnarray*}
\mbox{(\ref{equA})} &\leq &  (t+2)^{-t/2}a(t)\\
&&{}\cdot\sum_{i=1}^{k-t}
\frac{(k-t)(k-t-1)\cdots (k-t-i+1)}{%
n^{i/2}},
\end{eqnarray*}
which, for $b<1/2$, converges to $0$ as $n$ tends to infinity. A similar
lower bound for (\ref{equA}) shows that for $b\geq 1/2$ the sum cannot converge to
zero.

For the Bayes\vspace*{1pt} factor $B_{j0}^{\mathrm{Mix}}$ the proof of the posterior model
consistency is similar and hence omitted.

\item[4.]  We now\vspace*{1pt} prove that for $B_{j0}^{\mathrm{IP}}$ and $\pi ^{\mathrm{HU}}(M_{j})$ posterior
consistency holds for $b<1$. For large $n$ we have that
\begin{eqnarray*}
\mbox{(\ref{equA})} &\approx &  (t+2)^{-t/2}\\
&&{}\cdot\sum_{j=t+1}^{k}
\biggl( \frac{n}{j+2} \biggr) ^{-j/2}n^{t/2}{\pmatrix{k-t
\cr
j-t}}\frac{j!(k-j)!}{t!(k-t)!},
\end{eqnarray*}
which simplifies to
\[
\mbox{(\ref{equA})}\approx (t+2)^{-t/2}\sum_{j=t+1}^{k}
\biggl( \frac{n}{j+2} \biggr) ^{-j/2}n^{t/2}
\frac{j!}{(j-t)!}.
\]%
Making the change of variable $i=j-t$, the expression adopts the form
\[
\mbox{(\ref{equA})}\approx (t+2)^{-t/2}\sum_{i=1}^{k-t}
\frac{b_{i}}{n^{i/2}},
\]%
where
\[
b_{i}=(t+i+2)^{({t+i})/{2}}\frac{(t+i)!}{i!}.
\]%
Every individual term $b_{i}/n^{i/2}$ in the sum converges to~$0$ as $n$
tends to infinity, and for large values of $i$, the summands $b_{i}/n^{i/2}$
can be approximated by
\[
\frac{e^{t/2+1}\,i^{(i+3t)/2}}{n^{i/2}}.
\]%
For every $t$, this function of $i$ is decreasing for all $i<i_{0}$ and
increasing for $i>i_{0}$, where $i_{0}$ is given by
\[
i_{0}=-\frac{3t}{W ( -3et/n ) }\approx n/e-3t.
\]%
But as $k=O(n^{b})$ with $b<1$, this implies that the sequence $%
b_{i}/n^{i/2} $ is decreasing in $i$ for all $i\leq k$. Then, it follows
that the sum is upper bounded as
\[
\sum_{i=i_{0}+1}^{k-t}b_{i}/n^{i/2}
\leq \frac{b_{1}}{n^{1/2}}+(k-t)\frac{%
b_{2}}{n}.
\]%
For $k=O(n^{b})$ with $b<1$, the limit of the right-hand side of this
equation is $0$ when $n$ tends to infinity, and hence posterior model
consistency holds.

The proof of the consistency for the Bayes factor for the mixture of $g$-%
priors follows exactly the same pattern and it is therefore
omitted.

\item[5.]  For $b=1$, the proof of the posterior model consistency for $B_{j0}^{\mathrm{Mix}}$
and $\pi ^{\mathrm{HU}}(M_{j})$ runs as follows. For large $n$, it follows that,
under the alternative model $M_{t}$,
\begin{eqnarray*}
\frac{B_{j0}^{\mathrm{Mix}}}{B_{t0}^{\mathrm{Mix}}}
&\approx &  \frac{(({ne})/({j+1}))^{-j/2}}{(%
({ne})/({t+1}))^{-t/2}}\\
&&{}\cdot\frac{(1+\delta _{tj}^{\ast }-j/n)^{-(n-j-2)/2}}{%
(1-t/n)^{-(n-t-2)/2}}.
\end{eqnarray*}
The ratio $\pi ^{\mathrm{HU}}(M_{j})/\pi ^{\mathrm{HU}}(M_{t})$ of the model probabilities for
the hierarchical uniform prior is
\[
\frac{\pi ^{\mathrm{HU}}(M_{j})}{\pi ^{\mathrm{HU}}(M_{t})}=\frac{j!(k-j)!}{t!(k-t)!}.
\]%
Then, reasoning as before, for large $n$, the double sum~(\ref{equA}) of Theorem~\ref{THM1},
after some simplifications, can be approximated as
\begin{eqnarray*}
\mbox{(\ref{equA})}&\approx & (t+1)^{-t/2}\\
&&{}\cdot\sum_{j=t+1}^{k}
\biggl( \frac{ne}{j+1} \biggr) ^{-j/2}n^{t/2}\\
&&\hspace*{10pt}\qquad{}\cdot
\frac{j!}{(j-t)!} \biggl( 1-\frac{j}{n} \biggr) ^{-(n-j-2)/2}.
\end{eqnarray*}
Making the change of variable $i=j-t$, some further simplifications on the
factorials yield the approximating expression
\begin{eqnarray*}
\mbox{(\ref{equA})} &\approx &  \frac{e^{-3t/2}}{(t+1)^{t/2}}\sum_{i=1}^{k-t}
\frac{%
e^{-i/2}\,i^{-(i+1/2)}(i+t)^{(3i+3t)/2}}{n^{i/2}} \\
&&\hspace*{39pt}\qquad{}\cdot\biggl( 1-\frac{i+t%
}{n} \biggr) ^{-(n-i-t-2)/2}.
\end{eqnarray*}
Letting $x=i/k$ and $s=n/k$, the sum in the preceding expression can be
approximated, up to a constant, by the integral $\int_{0}^{1}f_{k}(x|s,t)%
\,dx $, where
\begin{eqnarray*}
f_{k}(x|s,t) &=&k(kx)^{-kx-({1}/{2})}(ks)^{-({kx})/{2}}\\
&&{}\cdot e^{-({1}/{2})kx}(kx+t)^{({3kx})/{2}+({3t})/{2}+{1}/{2}}
\\
&&{}\cdot \biggl( 1-\frac{kx+t}{ks} \biggr) ^{({1}/{2})(k(x-s)+t+2)}.
\end{eqnarray*}
We now prove that $\lim_{k\rightarrow \infty }$ $\int_{0}^{1}f_{k}(x|s,t)%
\,dx=0$ for any $t=0,1,2,\ldots$ and $s\geq 1$.

For any $k$, $t$ and $s\geq 1$, $f_{k}(x|s,t)>0$. For $t=0$, we have that $%
f_{k}(x|s,0)$ is a decreasing function of~$x$ for all $k$ and $s\geq 1$, and
such that $f_{k}(0|s,0)=k$. Further, $\lim_{k\rightarrow \infty
}f_{k}(x|s,0)=0$ for all $x\in (0,1]$. For $t=1,2,\ldots,$ even though $%
f_{k}(x|s,t)$ is not a decreasing function of $x$, except for large values
of $x$, we have that $\lim_{x\rightarrow 0}f_{k}(x|s,t)=0$, and $%
\lim_{k\rightarrow \infty }f_{k}(x|s,0)=0$ for all $x\in (0,1]$.

Thus, for any $t$, the limit of $f_{k}(x|s,t)$ when $k$ goes to infinity is
given by
\[
\lim_{k\rightarrow \infty }f_{k}(x|s,t)=
\cases{ \infty, &
$\mbox{if } x=0$,
\vspace*{2pt}\cr
0, & $\mbox{if } x\in (0,1]$,}
\]
and thus
\[
\int_{0}^{1}\lim_{k\rightarrow \infty }f_{k}(x|s,t)
\,dx=0.
\]

On the other hand, $f_{k}(x|s,t)$ is a decreasing function of $s$ and,
therefore, $f_{k}(x|s,t)\leq f_{k}(x|1,t)$. Moreover, for every
$t=0,1,2,\ldots$
there exists an integrable function $u(x|t)$, such that
\[
f_{k}(x|s,t)\leq u(x|t),
\]
for large values of $k$. For instance, the function $%
u(x| t)=10^{t}\operatorname{Ga}(x|0.1,1) $, where $\operatorname{Ga}(x|0.1,1)$ denotes the Gamma density
with parameters $0.1$ and $1$, is an upper bound of $f_{k}(x|s,t)$.

Applying the dominated convergence theorem to the sequence $\{f_{k}(x|s,t)$,
$k\geq 1\}$, we have that
\[
\lim_{k\rightarrow \infty }\int_{0}^{1}f_{k}(x|s,t)
\,dx=\int_{0}^{1}\lim f_{k}(x|s,t)
\,dx=0,
\]
and this completes the posterior model consistency proof for the Bayes
factor based on the mixture of $g$-priors and the hierarchical uniform prior.

A similar proof can be given for the Bayes factors for $g=n$ and for the
intrinsic priors. This completes the proof of Theorem~\ref{THM3}.
\end{longlist}

\end{appendix}

\section*{Acknowledgments}

We are very grateful to two anonymous referees for
very valuable comments that helped to improve the original manuscript. This
paper has been supported by grant MTM2011-28945, Ministerio de Ciencia,
Spain.






%
\end{document}